\documentstyle[emulateapj,epsf]{article}

\newcommand {\simlt}{\lower.5ex\hbox{$\; \buildrel < \over \sim \;$}}
\newcommand {\simgt}{\lower.5ex\hbox{$\; \buildrel > \over \sim \;$}}
\newcommand {\boom}{{\sc Boomerang}}

\newcommand {\boomna}{{\sc Boomerang}/NA}
\slugcomment{}

\begin{document}

\title {A measurement of $\Omega$ from the North American Test Flight of \boom}

\author{
A. Melchiorri\altaffilmark{1,2,9}, P.A.R. Ade\altaffilmark{3},
P. de Bernardis\altaffilmark{1}, J.J. Bock\altaffilmark{4,5}, 
J. Borrill\altaffilmark{6,7}, A. Boscaleri\altaffilmark{8}, B.P.
Crill\altaffilmark{4},
G. De Troia\altaffilmark{1}, P. Farese\altaffilmark{10}, P. G.
Ferreira\altaffilmark{9,11,12}, K. Ganga\altaffilmark{4,13}, G. de
Gasperis\altaffilmark{2}, 
M. Giacometti\altaffilmark{1}, 
V.V. Hristov\altaffilmark{4}, A. H. Jaffe\altaffilmark{6}, 
A.E. Lange\altaffilmark{4}, 
S. Masi\altaffilmark{1}, P.D. Mauskopf\altaffilmark{14},
 L. Miglio\altaffilmark{1,15}, 
C.B. Netterfield\altaffilmark{15}, 
E. Pascale\altaffilmark{8}, F. Piacentini\altaffilmark{1},
G. Romeo\altaffilmark{16}, J.E. Ruhl\altaffilmark{10} and N.
Vittorio\altaffilmark{2}
}

\affil{ 
 $^1$ Dipartimento di Fisica, Universita' La Sapienza, Roma, Italy \\
 $^2$ Dipartimento di Fisica, Universita' Tor Vergata, Roma, Italy \\
 $^3$ Queen Mary and Westfield College, London, UK \\
 $^4$ California Institute of Technology, Pasadena, CA, USA \\
 $^5$ Jet Propulsion Laboratory, Pasadena, CA, USA \\
 $^6$ Center for Particle Astrophysics, University of California,
Berkeley, CA, USA \\
 $^7$ National Energy Research Scientific Computing Center, LBNL,
Berkeley, CA, USA \\
 $^8$ IROE-CNR, Firenze, Italy \\
 $^9$ Dept. de Physique Theorique, Universite de Geneve, Switzerland\\
 $^{10}$ Dept. of Physics, Univ. of California, Santa Barbara, CA, USA \\
 $^{11}$ CENTRA, IST, Lisbon, Portugal \\
 $^{12}$ Theory Division, CERN, Geneva, Switzerland \\
 $^{13}$ Physique Corpusculaire et Cosmologie, College de France, 11 place
Marcelin Berthelot, 75231 Paris Cedex 05, France\\
 $^{14}$ Dept. of Physics and Astronomy, University of Massachussets, Amherst, MA, USA \\
 $^{15}$ Depts. of Physics and Astronomy, University of Toronto, Canada
\\
 $^{16}$ Istituto Nazionale di Geofisica, Roma, Italy \\
}

\begin{abstract}

We use the angular power spectrum of the Cosmic Microwave Background,
measured during the North American test flight of the {\boom } experiment,
to constrain the geometry of the universe. Within the class
of Cold Dark Matter models, we find 
that the overall fractional energy density of the universe, $\Omega$, is
constrained to be $0.85 \le \Omega \le 1.25$ at the $68\%$ confidence level.
Combined with the COBE measurement and the high redshift supernovae data 
we obtain new constraints on the fractional matter density 
 and the cosmological constant.

\end{abstract}


\section{Introduction}
The dramatic improvement in the quality of astronomical data
in the past few years has presented cosmologists with the
possibility of measuring the large scale properties
of our universe with unprecedented precision (e.g. \cite{review}).
The sensitivity of the angular power spectrum of
the Cosmic Microwave Background (CMB) to cosmological parameters
has lead to analyses of existing datasets
with increasing sophistication in an attempt to measure
such fundamental quantities as the energy density of
the universe and the cosmological constant. This
activity has lead to improved methods of Maximum 
Likelihood Estimation (Bond, Jaffe \& Knox 1998, 
hereafter \cite{BJK98}, \cite{bartlett}), attempts at enlarging the
range of possible parameters (\cite{lineweaver,teg98,melk99}),
and the incorporation of systematic uncertainties in 
the experiments (Dodelson \& Knox,  from now on \cite{dk99},
\cite{ganga}).

Within the class of adiabatic inflationary models
 there is now strong evidence
from the CMB that the universe is flat. 
The most extensive range of parameters has been
considered by Tegmark (1999) where the author found
that a flat universe was consistent with CMB data 
at the $68 \%$ confidence level. A more thorough analysis
was performed in \cite{dk99}, incorporating the non-Gaussianity
of the likelihood function, possible calibration 
uncertainties and the most recent data: again, the
$68 \%$ likelihood contours comfortably encompass
the Einsten-de Sitter Universe. 
All these previous analyses were restricted to
the class of open and flat models.

In this {\it letter} we present further evidence for
a flat universe from the CMB. Using the methods for
parameter estimation described in \cite{BJK98}, we 
perform a search in cosmological parameter space
for the allowed range of values for the fractional density
of matter, $\Omega_M$, and cosmological constant, $\Omega_\Lambda$,
given the recent estimate of the angular power spectrum from
the 1997 test flight of the {\boom } experiment
 (see the companion paper \cite{mausk99}).
We obtain our primary constraints from this data set {\it alone}
and find compelling evidence, {\it within the family of 
adiabatic inflationary models}
for a flat universe. In section \ref{data}
we briefly describe the {\boom } experiment, the 
data analysis undertaken and the characteristics of
the angular power spectrum obtained. In section \ref{method}
we spell out the parameter space we have explored, the
method we use and the constraints we obtain on the
fractional energy density of the universe, 
$\Omega=\Omega_M+\Omega_\Lambda$. Finally in section
\ref{discussion} we discuss our findings and combine it with
other cosmological data to obtain a 
new constraint on the cosmological constant.

\newpage

\section{The Data}
\label{data}

The data we use here are from a North American test flight of
{\boom } (\boomna), a balloon-borne telescope designed to map CMB anisotropies
from a long-duration, balloon-borne (LDB) flight above the Antarctic.
 A detailed description of
the instrument can be found in \cite{masi99}.  A description of
the data and observations, with a discussion of calibrations, systematic
effects and signal reconstruction can be found in \cite{mausk99}.
This test flight produced maps of the CMB with more than 200 square
degrees of sky coverage at frequencies of 90 and 150~GHz with
resolutions of 26 arcmins FWHM and 16.6 arcmins FWHM respectively.

The size of the {\boomna } 150~GHz map (23,561, $6'$ pixels)
required new methods of analysis able to incorporate the effects of
correlated noise and new implementations capable of processing large
data sets. The pixelized map and angular power spectrum were produced
using the MADCAP software package of Borrill (1999a, 1999b) (see
http://cfpa.berkeley.edu/$\sim$borrill/cmb/madcap.html) on the Cray T3E-900
at NERSC and the Cray T3E-1200 at CINECA.

\medskip 
\vbox{\epsfxsize=8.0cm\epsfbox{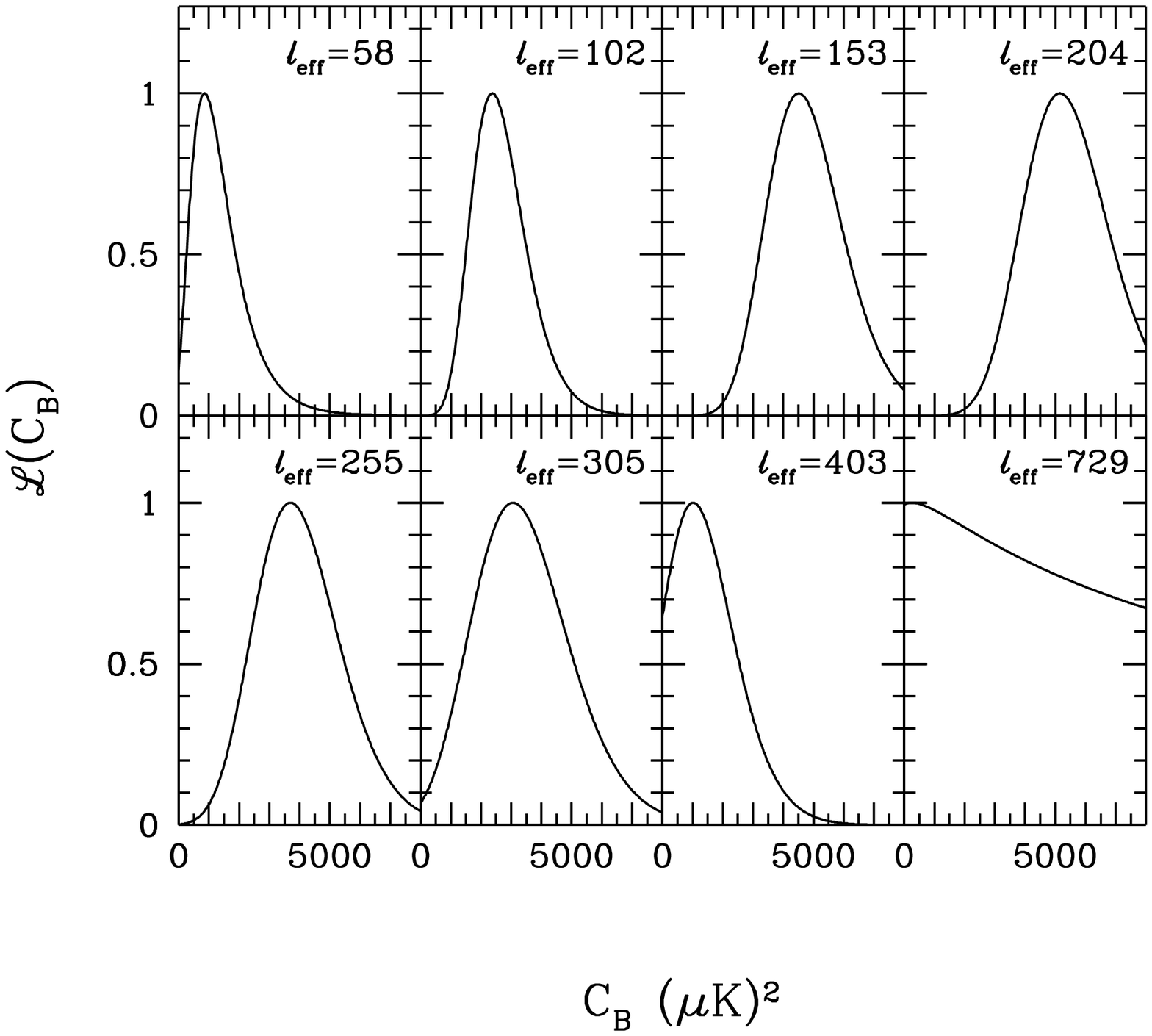}}
{
\small F{\scriptsize IG}.~1.--- 
The Likelihood function of each of the eight band powers, $\ell_{\rm eff} = 
(58, 102, 153, 204, 255, 305, 403, 729)$, reported in
\cite{mausk99} computed using the offset lognormal ansatz of \cite{BJK98}.
\label{fig:bandpowers}} \medskip

The angular power spectrum, $C_\ell$, resulting from the analysis of the
150~GHz map was estimated in eight bins spanning $\ell$ with
 seven bin's centered between $\ell = 50$ to $\ell = 400$ and one
bin at $\ell = 800$.


The bin correlation matrix
is diagonalized as in BJK98 resulting in eight orthogonalized
(independent) bins. 
We present the likelihood for each orthogonalized band
power in Figure 1, using the offset lognormal ansatz proposed
in \cite{BJK98}. 
As described in \cite{mausk99} the data
show strong evidence for an acoustic peak with an amplitude
of $\sim70\mu$K$_{CMB}$ centered at $\ell\sim200$.

 \section{Measuring Curvature}
\label{method}
The {\boomna } angular power spectrum covers
a range of $\ell$ corresponding to the horizon size at
decoupling. The amplitude and shape
of the power spectrum is primarily sensitive to the overall curvature
of the universe, $\Omega$ (\cite{dsz78}); other parameters such as the
scalar spectral index, $n_S$, the fractional energy density in
baryons, $\Omega_B$, the cosmological constant, $\Omega_\Lambda$, and
the Hubble constant, $H_0\equiv 100h$ km sec$^{-1}$, will also affect the
height of the peak and therefore some ``cosmic confusion'' will arise if
we attempt individual constraints on each of the parameters
(\cite{bond94}).
In our analysis we shall restrict ourselves to the family of 
adiabatic, CDM models. This involves considerable theoretical
predjudice in the set of parameters we choose to vary although,
as the presence of an acoustic peak at $\ell\sim200$ becomes
more certain, the assumption that structure was seeded by
primordial adiabatic perturbations becomes more compelling
(\cite{liddle}; however, counterexamples exist,
\cite{turok,durrer,hu}).

We should, in principle, consider an 11-dimensional space of 
parameters; sensible priors due to previous constraints
and the spectral coverage of the {\boomna } angular power spectrum
reduce the space to 6 dimensions.
In particular, we assume $\tau_c=0$ (lacking convincing evidence for
high redshift reionization), we assume a negligeable contribution of
gravitational waves (as predicted in the {\it standard} scenario),
 and we discard the weak effect due to massive neutrinos.
The remaining parameters to vary are $\Omega_{CDM}$,  $\Omega_\Lambda$, 
$\Omega_B$, $h$, $n_S$ and the amplitude of
fluctuations, $C_{10}$, in units of $C_{10}^{COBE}$.
 The combination $\Omega_Bh^2$ is constrained
by primordial nucleosynthesis arguments: {$0.013\le \Omega_Bh^2 \le 0.025$},
while we set $0.5 \le h \le 0.8$. For the spectral index of the primordial
scalar fluctuations we make the choice $0.8 \le n_S \le 1.3$ and we let 
a $20 \%$ variation in $C_{10}$. 
As our main goal is to obtain constraints
in the ($\Omega_M=\Omega_{CDM}+\Omega_B$, $\Omega_\Lambda$) plane, we let these
parameters vary in the range $[0.05,2] \times [0,1]$.
Proceeding as in \cite{dk99}, we attribute a likelihood to a point on this
plane by finding the remaining four, ``nuisance'', parameters that
maximize it. The reasons for applying this method are twofold. First,
if the likelihood were a multivariate Gaussian in all the parameters,
maximizing with regards to the nuisance parameters corresponds
to marginalizing over them. Second, if we define our $68\%$, $95\%$
 and  $99\%$ contours where the likelihood falls to $0.32$, $0.05$ and
$0.01$ of its peak value (as would be the case for a
two dimensional Multivariate Gaussian), then the constraints we obtain are
conservative relative to any other hypersurface we may choose
in parameter space in the sense that they rule out a smaller range of
parameter space than other usual choices.

The likelihood function for the estimated band powers is non-Gaussian
but one can apply the ``radical compression'' method proposed by
\cite{BJK98}; the likelihood function is well approximated by
an offset lognormal distribution whose parameters can be
easily calculated from the output of MADCAP. The theory $C_\ell$s
are generated using CMBFAST (\cite{sz}) and the recent implementation
for closed models CAMB (\cite{levin}). We search for the maximum
along a 4 dimensional grid of models, using the fact that variations in 
$C_{10}$ and $n_s$ are less CPU time consuming.
We searched also for the multidimensional maxima of the likelihood adopting
a {\it Downhill Simplex Method} (\cite{nr}), obtaining consistent results.   

\centerline{\vbox{\epsfxsize=8.5cm\epsfbox{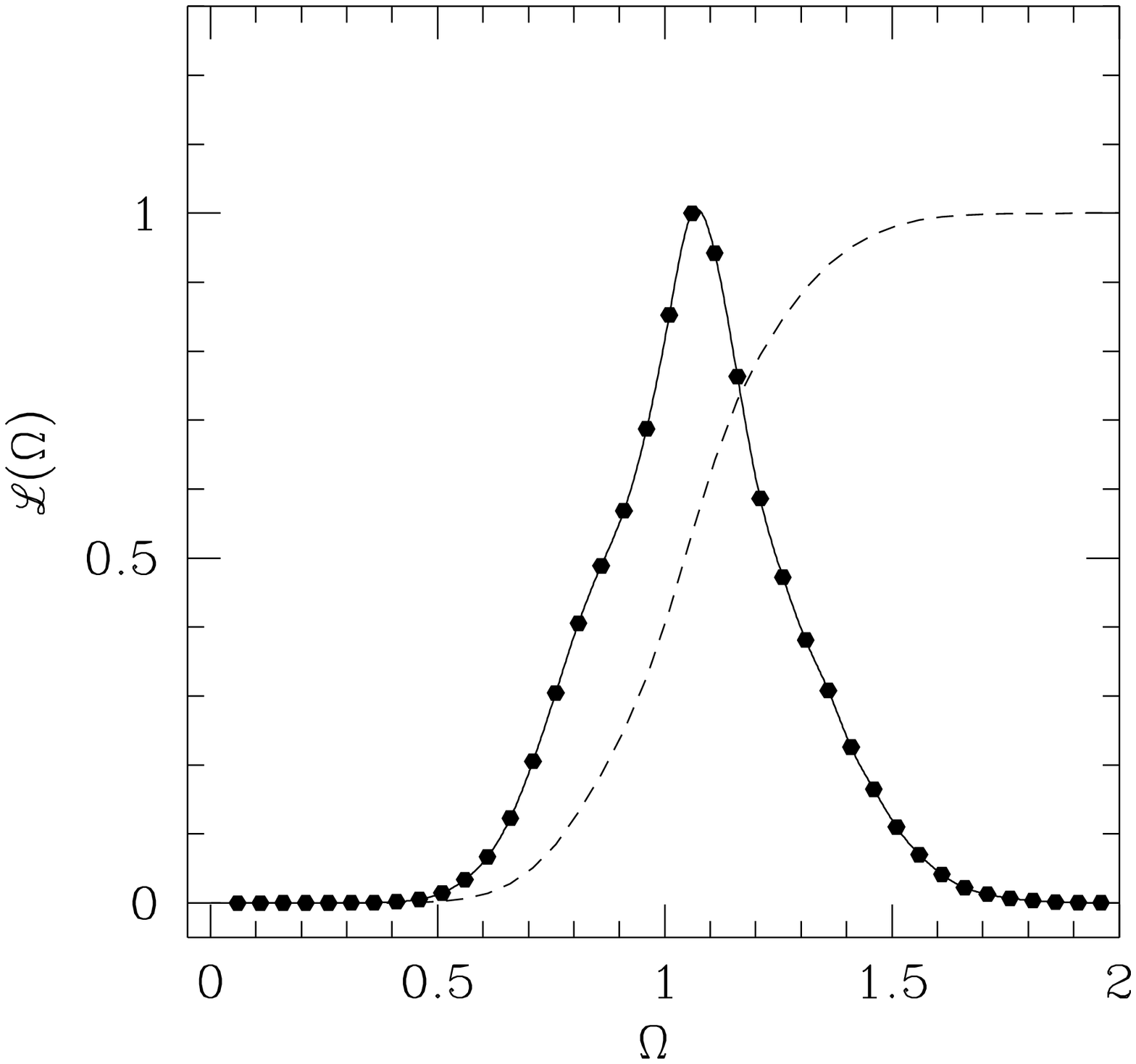}}}
{
\small F{\scriptsize IG}.~2.--- 
The Likelihood function of $\Omega= \Omega_M +\Omega_{\Lambda}$ normalized to unity at the peak
after  marginalizing along the 
$\Omega_M-\Omega_\Lambda$ direction. The dashed line is the cumulative
likelihood.
\label{fig:curvature}}
\vskip 1.5cm
 
In Figure 2, we plot the likelihood of $\Omega$ normalized to 1 at
the peak where, again, we have maximized along the
$\Omega_M-\Omega_\Lambda$ direction.  The
likelihood shows a sharp peak near $\Omega=1$ and this result
is insensitive to the tradeoff between $\Omega_M$ and 
$\Omega_{\Lambda}$.
(see Figure 3 and explanation in following paragraphs).
This is an extreme manifestation of the ``cosmic degeneracy'' problem
(because we are focusing on just the first peak): we are able
to obtain robust constraints on $\Omega$ without strong
constraints on $\Omega_M$ and $\Omega_\Lambda$ individually.

Within the range of models we are considering, 
we find that $68\%$ of integrated likelihood corresponds 
to $0.85 \le \Omega \le 1.25 $ ($0.65 \le \Omega \le 1.45$ at $95 \%$).
The best fit is a marginally closed model with 
$\Omega_{CDM}=0.26$, $\Omega_{B}=0.05$, $\Omega_{\Lambda} =0.75$, 
$n_{S}=0.95$, $h = 0.70$, $C_{10}=0.9$.
An almost equivalent good fit is given by $\Omega_{CDM}=0.39$, 
$\Omega_{B}=0.07$, $\Omega_{\Lambda} =0.65$, $n_{S}=0.90$, $h = 0.55$,
 $C_{10}=1.0$.

In Figure 3 (top panel) we estimate the likelihood of the data
for a $20\times20$ grid in ($\Omega_M$,$\Omega_\Lambda$) 
by applying the 
maximization/marginalization algorithm described above.
The effect of marginalizing is, as expected, to expand the contours along
the $\Omega=$constant lines but has little effect in
the perpendicular direction and we are able to rule out
a substantial region of parameter space.

For $\Omega_M \sim 1$ models, the position of the peak is
solely dependent on the angular-diameter distance, with
a good approximation being 
$\ell_{peak}\propto\Omega^{-\frac{1}{2}}$; this approximation
breaks down when $\Omega_M\rightarrow0$ 
where the early time integrated Sachs-Wolfe effect
becomes important and $\ell_{peak}$
is far more sensitive to $\Omega$ (\cite{WS96}). 
 This effect leads to a 
convergence of contour levels as $\Omega_M\rightarrow0$
in Figure 3.

\vbox{\epsfxsize=8.0cm\epsfbox{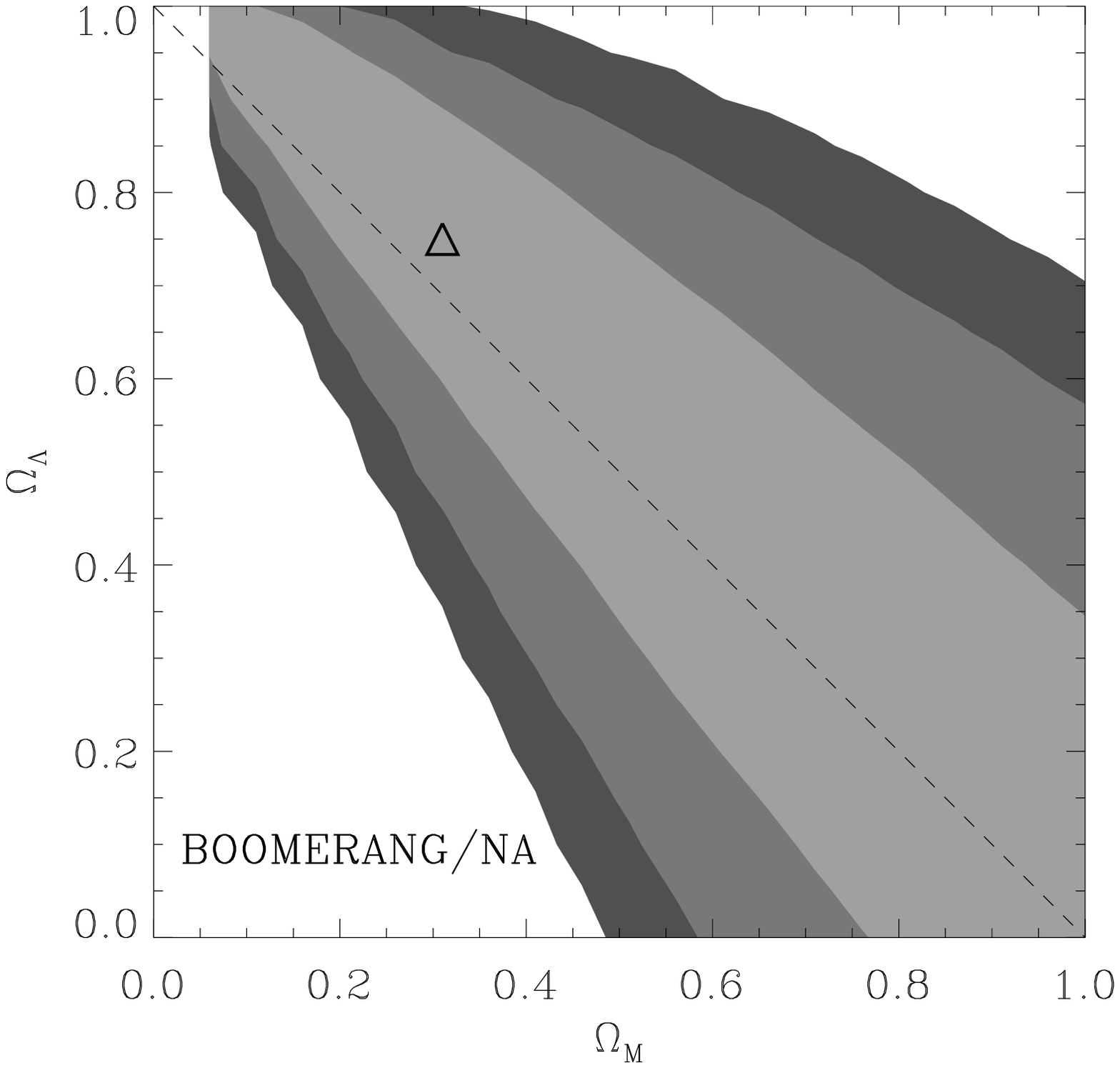}}
\vbox{\epsfxsize=8.0cm\epsfbox{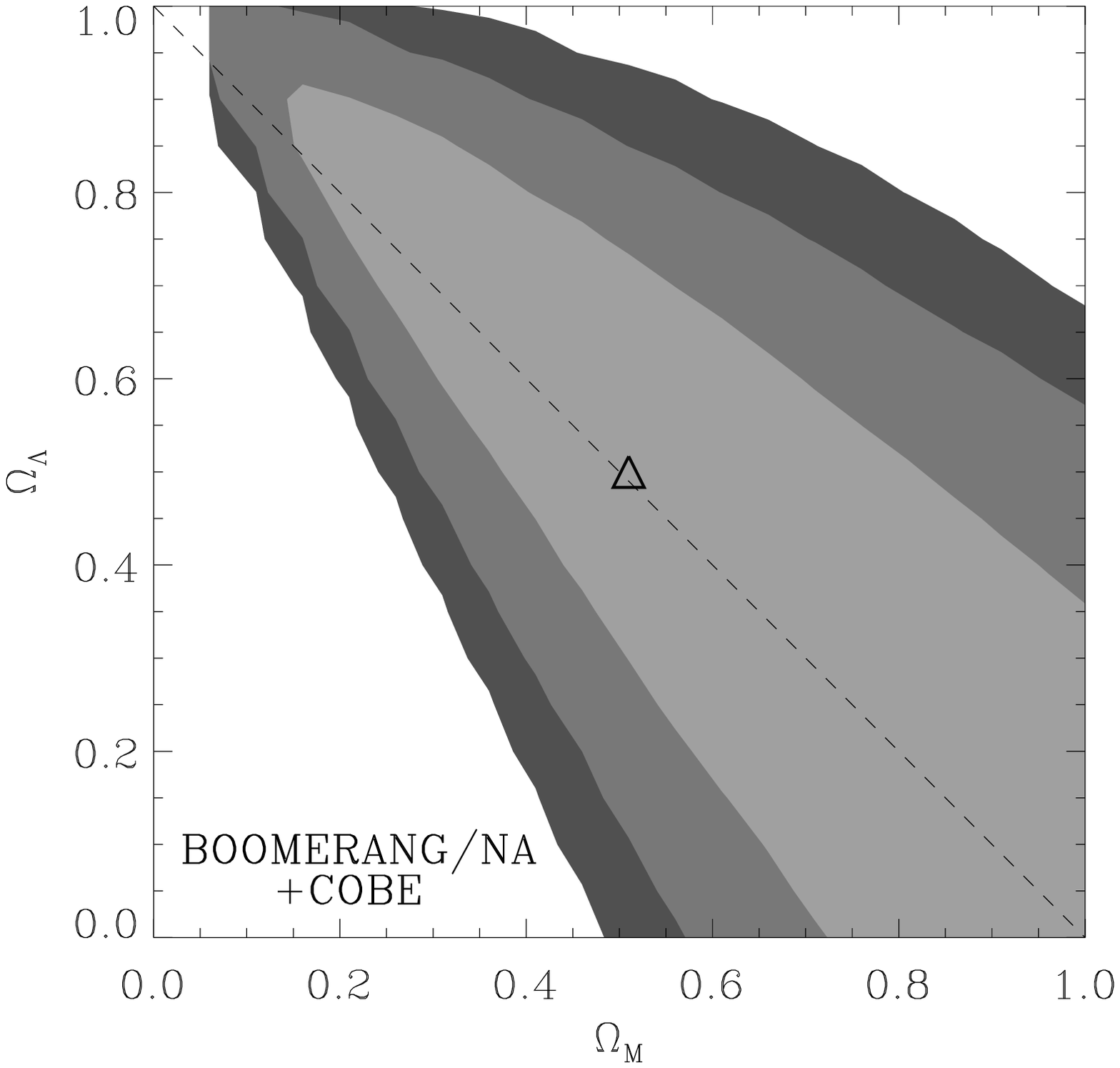}}
{\small F{\scriptsize IG}.~3.--- 
The likelihood contours in the $\Omega_M$,$\Omega_\Lambda$ plane,
evaluated at the maxima of the remaining four ``nuisance'' parametes.
The top panel is from the {\boomna } data, the bottom panel 
is from \boomna+COBE. The contours correspond to  
0.32, 0.05 and 0.01 of the peak value of the
likelihood. The small triangle indicates the best fit. 
The dashed line corresponds to the flat models.
\label{fig:like_mar}}
\vskip 1cm

\section{Discussion}

In the previous section we have obtained a
constraint on $\Omega$ using only the {\boomna } data. These
new results are consistent with Lineweaver (1998), Tegmark (1999)
and \cite{dk99}.  However, the {\boomna } data on its own does not constrain
the shape and amplitude of the power spectrum at $\ell \le 25$ and limits
our ability to independently determine the parameters
$n_S$, $\Omega_B$, $h$, $\Omega_\Lambda$ and $C_{10}$. We combine
the {\boomna } data with the 4-year COBE/DMR angular power spectrum
to attempt to break this degeneracy. 
In Figure 3 (bottom panel) we plot the likelihood contours, again maximized
over the nuisance parameters for the combined {\boomna } and COBE data.
The inclusion of the COBE data does not greatly affect the
constraints at high $\Omega_M$ or the confidence levels on $\Omega$,
but, as expected, it helps to close of the contours at low values of
$\Omega_M$.  The best fit model changes to have
$\Omega_{CDM}=0.46$, $\Omega_{B}=0.05$, $\Omega_{\Lambda} =0.50$, 
$n_{S}=1.0$, $h = 0.70$, $C_{10}=0.94$. We find that for the likelihood 
to be greater than $0.32$ of its peak value then $\Omega_M>0.2$, 
again similar to the results of \cite{dk99}.

 One can combine our constraints with those 
obtained from the luminosity-distance measurements of
high-$z$ supernovae (\cite{SN1a,SN1b}): 
using the 1-$\sigma$ constraint from Perlmutter et al (1998), 
$\Omega_M - 0.6 \Omega_\Lambda = -0.2\pm 0.1$, we find
$0.2\le\Omega_M\le0.45$ and $0.6\le\Omega_\Lambda\le0.85$.


A few comments are in order about the robustness of
our analysis.
Firstly we have not truly marginalized over
the nuisance parameters. However the constraints 
we obtain in this way are, if anything, more conservative.
Secondly, although we
are limiting ourselves to standard adiabatic models,
a strong case can be made against the rival theory
of topological defects: the presence of a fairly localized
rise and fall in the data around $\ell$ of 200
indicates that the characteristic broadening due
to decoherence of the  either cosmic strings (\cite{chm98}) 
or textures (\cite{pst98}) is strongly disfavoured.

Finally we have restricted ourselves to only four extra
nuisance parameters. Again we believe this does
not affect our main result (our constraints on $\Omega$)
although it may affect the low $\Omega_M$ constraints
when we combine the {\boomna } data with COBE; the results
from Tegmark (1998) and \cite{dk99} lead us to believe
that the effect will not greatly change our results.

To summarize we have used the angular power spectrum of the
{\boomna } test flight to constrain the curvature of the universe.
Given that we have based our results on this data set alone,
our results are completeley independent from previous analysis
of the CMB. At the time of submission, this {\it letter}
 is also the first analysis of this kind to include
closed models in the computation.

We find strong evidence against an open universe:
we find that $0.65 \le \Omega \le 1.45$  at the 95$\%$
confidence level, significantly ruling out the current
favourite open inflationary models for structure formation
(\cite{ls,rp,bt}).
Much tighter constraints will soon be placed on these and others cosmological
parameters from future data sets, including data obtained during 
by the Antarctic LDB flight of \boom, which mapped over $1200$ square 
degrees of the sky with $12'$ angular resolution and higher sensitivity
per pixel than the {\boomna}.

\label{discussion}
 
\acknowledgments 
We acknowledge useful conversations with Dick Bond, Ruth Durrer, 
Eric Hivon, Tom Montroy, Dmitry Pogosyan and Simon Prunet. 
The \boom program has been supported by Programma Nazionale Ricerche in Antartide, Agenzia Spaziale Italiana and University of Rome La Sapienza
in Italy, by NASA grant numbers NAG5-4081 \& NAG5-4455 in the USA, and 
by the NSF Science \& Technology
Center for Particle Astrophysics grant number SA1477-22311NM under
AST-9120005, by the NSF Office of Polar Programs grant number OPP-9729121
and by PPARC in UK.
This research also used resources of the National Energy Research
Scientific Computing Center, which is supported by the Office of
Science of the U.S. Department of Energy under Contract
No. DE-AC03-76SF00098.  Additional computational support for the
data analysis has been provided by CINECA/Bologna.
We also acknowledge using the CMBFAST, CAMB and RADPack packages.


\end{document}